\documentclass[mathleft
]{an}
\usepackage{graphicx}
\usepackage{times}
\begin{document}

\Pagespan{789}{}
\Yearpublication{2008}%
\Yearsubmission{2008}%
\Month{11}%
\Volume{999}%
\Issue{88}%

\title{Radial velocities, dynamics of stars and nebulosities with GAIA and VLT-GIRAFFE}

\author{C. Martayan\inst{1,2}\fnmsep\thanks{Corresponding author:
  \email{martayan@oma.be}\newline}
\and Y. Fr\'emat \inst{1}
\and R. Blomme \inst{1}
\and A. Jonckheere \inst{1}
\and C. Delle-Luche \inst{2}
\and P. Sartoretti \inst{2}
\and D. Katz \inst{2}
\and Y. Viala \inst{2}
\and M. Floquet \inst{2}
\and A.-M. Hubert \inst{2}
\and C. Neiner \inst{2}
}
\titlerunning{Radial velocities with GAIA and VLT-GIRAFFE}
\authorrunning{C. Martayan et al.}
\institute{
Royal Observatory of Belgium, 3 avenue circulaire, 1180 Brussels, Belgium
\and 
GEPI, Observatoire de Paris, CNRS, Universit\'e Paris Diderot; place Jules Janssen 92195 Meudon Cedex, France
}

\received{}
\accepted{}
\publonline{later}


\abstract{%
This document is divided in two parts. The first part deals with the radial velocities (RV) distributions for
B-type stars and nebulosities observed with the VLT-GIRAFFE in the Large and Small Magellanic Clouds towards the
open clusters NGC2004 and NGC330. Thanks to the resolution of GIRAFFE spectra, we found that the RV distribution
for the nebulosities in the LMC is bi-modal. This bi-modality can be interpreted, in term of dynamics, by the
expansion of the LMC4 superbubble. The second part deals with the GAIA space mission and the
determination of the radial velocities by using Radial Velocity Spectrometer (RVS) spectra. The methods to 
determine the radial velocities are presented as well as preliminary results on simulated RVS spectra.
}

\maketitle


\section{Radial velocity distributions of stars and nebulosities with the VLT-GIRAFFE}

We observed large samples of B-type stars in the Large and Small Magellanic Clouds towards 
the open clusters \\
NGC2004 in the LMC and NGC330 in the SMC. 

Thanks to the resolution of GIRAFFE spectra \\
(R= 6400 for LR02 spectra 396.4 -- 456.7nm,
R= 8600 for LR06 spectra 643.8 -- 718.4nm), we were able to distinguish the Halpha emission lines, which come from the
circumstellar disks of Be stars (broad emission), 
from the nebular emission lines (narrow emission). 
Moreover, we were able to see the other nebular lines of [NII] 654.8-658.3 nm and [SII]
671.7-673.1nm. We determined the radial velocities of the nebulosities from these lines.

For the stars, we measured the radial velocities by fitting the LR02 observations with theoretical spectra by using the
GIRFIT code (Fr\'emat et al. 2006). This code provides the fundamental parameters of the stars (Teff, logg, Vsini, and
the radial velocity).  The accuracy of the determination of the radial velocities is better than 10 km/s.

We then compared the distributions of radial velocities of the stars with the distributions of radial velocities for
the nebular lines. In the SMC, the 2 distributions are gaussian and peak at the same value: $\sim$160 km/s indicating a
possible link between the stars and the nebulosities. However, in the LMC, the RV distribution for the stars is gaussian, 
mono-modal, and peaks at 300 km/s; while the RV distribution for the
nebular lines is clearly bi-modal and peaks at 305 and 335 km/s. No clear link is found between the stellar object and
the nebular emission in the spectrum.

The high and low nebular RVs are not randomly distributed over the field
but seem to be organized in structures that look like filaments. The higher velocities centred around 335 km/s are
observed in the eastern and north-western parts of the field, while the lower velocities centred around 305 km/s, are
observed in the south-western part. Higher intensity is generally found in nebulosities with the lowest H$\alpha$ radial
velocities. However, stars with no nebular H$\alpha$ line are found in isolated regions of the observed field, reflecting the
patchy nature of nebulosities in this part of the LMC.

The mean value of [SII] 6717/6731 is 1.4 typical of LMC bubbles (Skelton et al. 1999). The [NII]/H$\alpha$ is lower than 0.1,
the [SII]/H$\alpha$ ranges from 0.1 to 0.3. These values are close to those found for the HII regions in the LMC. The analysis
of nebular lines agrees well with the HI distribution survey of Staveley-Smith et al. (2003). The HI distribution
revealed that the body of the LMC is punctuated by ``large holes''. One of the main HI gaps in the LMC corresponds to LMC4.
The southern inner limit of the LMC4 supergiant shell crosses the field observed with the VLT-GIRAFFE spectrograph. The
LMC4 HI supergiant shell is centred at 05h31mn33s -66$^{o}$40'28'' with a radius of 38.7' and a systemic heliocentric
velocity of 306 km/s.

\begin{figure}
\includegraphics[width=83mm,height=70mm]{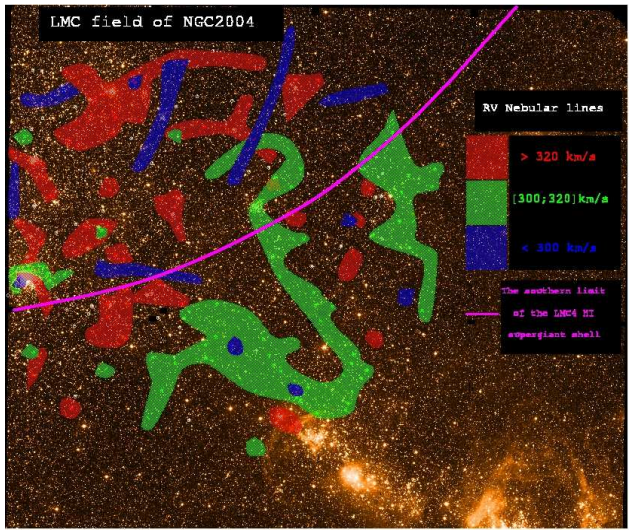}
\caption{Regions of similar radial velocities of the nebular lines. Red and blue regions are associated to the LMC4 bubble
(respectively with RVs $>$320 km/s and $<$300 km/s, which correspond to the expansion of the bubble). 
The pink curve corresponds to the southern limit of the LMC4 bubble.
The green regions are associated to the mean RV of the LMC
(300--320 km/s). This figure shows the patchy and filamentary nature of the nebulosity. 
North is at the top, East on the left.}
\label{fig1}
\end{figure}

Thus our study of nebular lines shows that:
\begin{itemize}
\item higher velocity nebulosities (335 km/s) appear to coincide with the southern inner HI gas in LMC4.
\item lower velocity nebulosities (305 km/s) are preferentially detected at the rim of LMC4 and have the same RV as 
the systemic RV of the bubble.
\item The separation between lower and higher velocities 
corresponds to the inner limit of the LMC4 HI supergiant shell.
\item The strong intensity of nebular lines observed in the south-western part of the field is easily explained by the 	  vicinity of the HII region LHA 120-N 15A. 
\end{itemize}
Fig.~\ref{fig1} shows the links between the different regions of the same radial
velocities for nebular lines. All the details about this study can be retrieved from Martayan et al. (2006).

\section{The GAIA space mission and the radial velocity determinations}

GAIA will observe 1 billion of stars both in the Milky Way and close galaxies of the Local Group (e.g. LMC, SMC for example).
It has 3 instruments on-board:  
\begin{itemize}
\item ASTRO for the determination of the astrometry, proper motions, and parallaxes of the stars.
\item Blue and red spectrophotometers (Bp/Rp) are 2 spectrophotometers, which will provide very low resolution spectra 
(R$\sim$100,320-680nm and 640-1000nm).  
\item The Radial Velocity Spectrometer (RVS) is the spectrograph of GAIA. It will
provide spectra at high resolution (R=11500) for stars till V$\sim$11, and low resolution spectra (R=5000) 
for stars till V$\sim$17.
The spectra will be used to determine the radial velocities of the stars.
\end{itemize}

\begin{figure*}[h!]
\begin{tabular}{lll}
\includegraphics[width=52mm,height=110mm]{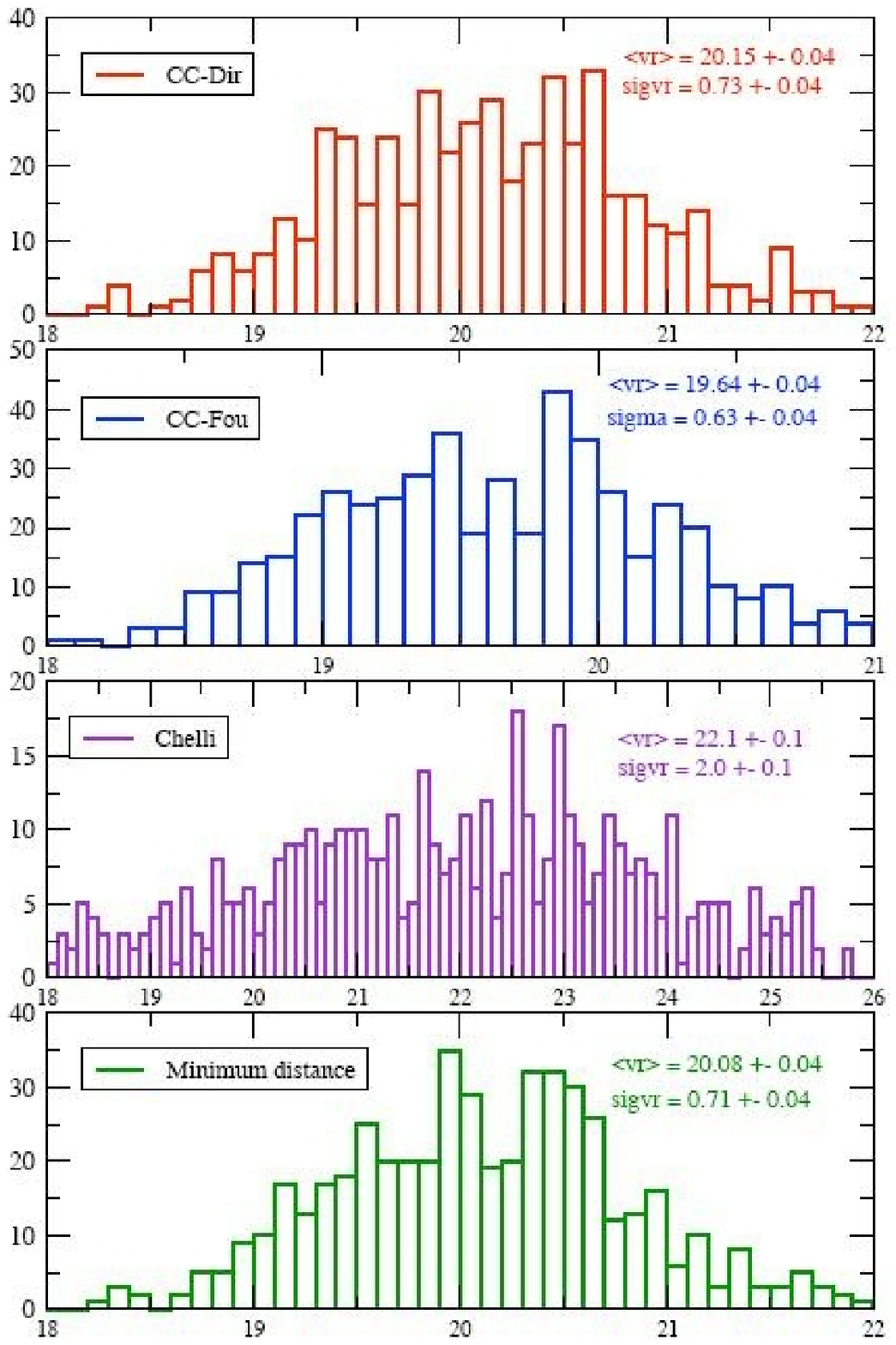} & \includegraphics[width=52mm,height=110mm]{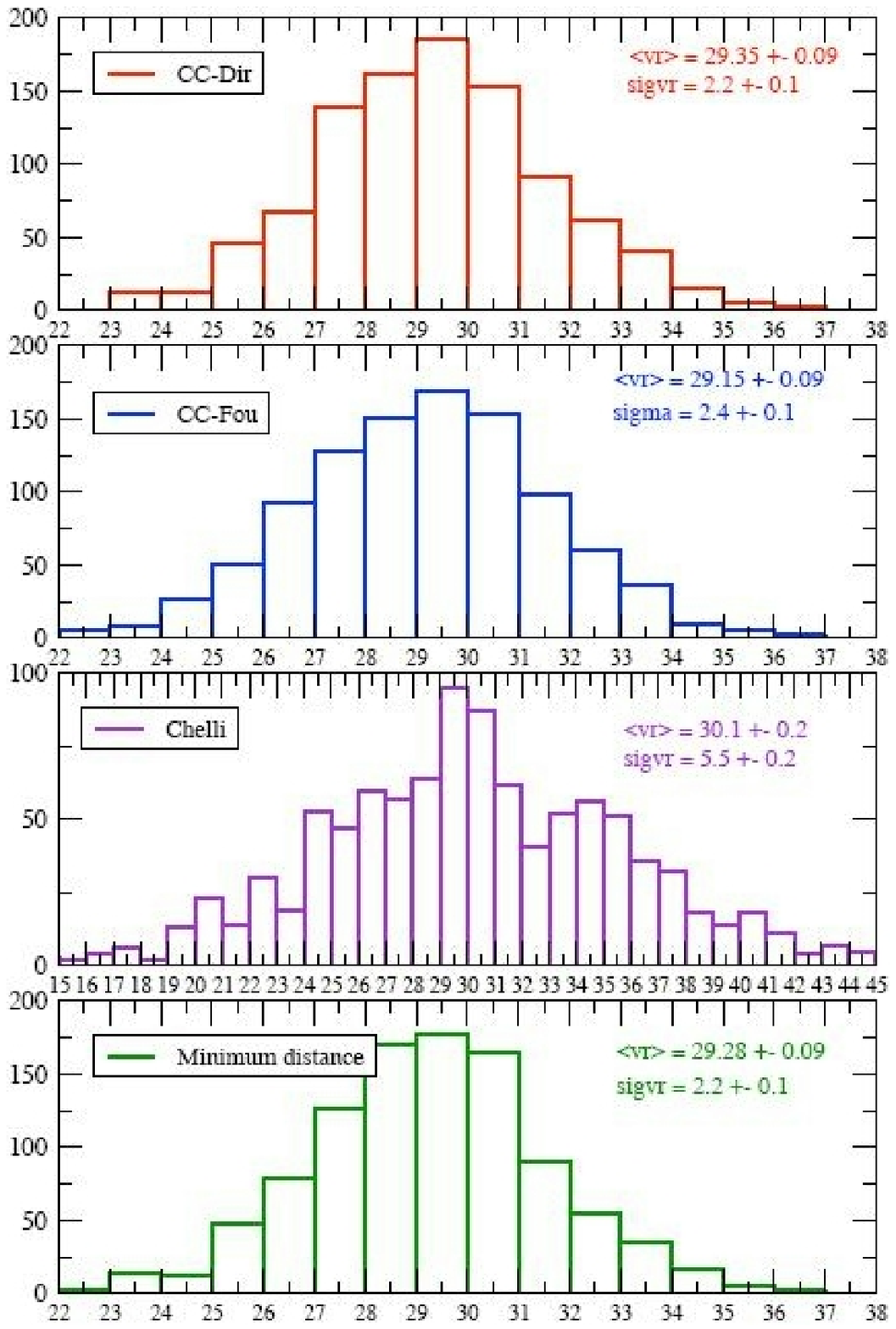} &
\includegraphics[width=52mm,height=110mm]{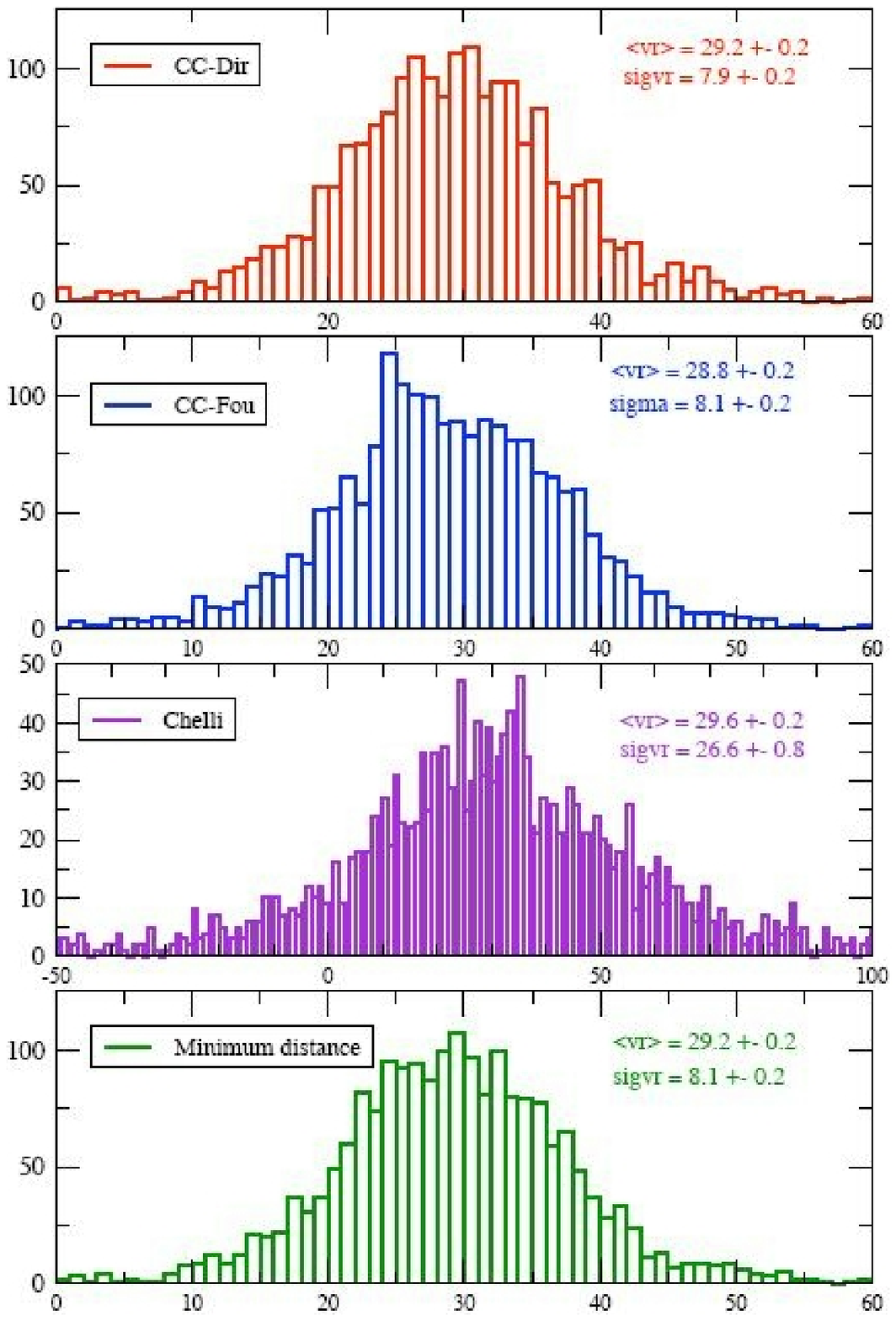}
\end{tabular}
\caption{ Histograms of the radial velocity distributions for a G5V star (Teff: 5500 K, logg=4.0, [Fe/H]=0.0, Vsini=0 km/s ) 
Left: magnitudeV=9.5 (500 high resolution spectra, RV=20 km/s), middle: V= 11.5 (1000 low resolution spectra, RV=30 km/s), 
right: V=13.5 (1000 low resolution spectra, RV=30 km/s). From top to bottom:  In red: with the method of cross-correlation 
in direct space (CC-Dir), in blue: with the cross-correlation in Fourier space (CC-Fou), 
in purple: with the Chelli's method in Fourier space (Chelli), in green: with the minimum distance method (MD).
X-axis corresponds to the RV (unit km/s), Y-axis to the number of spectra.
In each histogram, the average of the RV determination is indicated.}
\label{fig2}
\end{figure*}

Over 5 years, all the objects will be observed 40 times in average by the GAIA-RVS. The radial velocities will be
estimated from the single transit and from the combined transits. We present here the methods we currently implement in
java for the automatic determination of the radial velocities of the stars. We show the first tests and results of the
different methods on simulated data in Fig.~\ref{fig2}.

\subsection{Methods used for the determination of the stellar radial velocities}
For more details about the methods and the codes used, we refer the reader to the technical notes from 
Viala et al. (2007, 2008) available via the GAIA website or on demand.
 
\subsubsection{Determination of radial velocity through cross-correlation of an object spectrum 
with a template spectrum (or a mask) in direct space (CC-Dir)}
The objective of this scientific module is, 
by cross- correlation between the calibrated object spectrum and an appropriate 
template spectrum, to determine simultaneously the radial velocity and the projected rotational velocity of the object during 
a single transit.  It gets the appropriate template spectrum from the workpackage ``selection of synthetic spectra''. It shifts 
the template spectrum by applying a radial velocity and computes the cross-correlation coefficient, in data space, between 
the object spectrum and the shifted template. The radial velocity is determined from the maximum of the cross-correlation 
function, maximum obtained by fitting a parabola in the top of the correlation peak.

\subsubsection{Radial velocity determination through cross-correlation in Fourier space (CC-FOU and Chelli)}
The aim of this module is to compute the Cross-Correlation function of an object spectrum by a template in Fourier Space. 
For this method, we considered therefore two possibilities to estimate the radial velocities: 
first, we compute the Fourier inverse of the Fourier Cross-Spectrum in order to locate the peak value 
of the function (like in the CC-Dir approach) by accounting for pixel-discretisation; 
and second, we directly derive the radial velocity in Fourier space by adopting the approach described by Chelli (2000). 

\subsubsection{Determination of radial velocity by the minimum distance method (MD)}
Another method of determining the radial velocity is by minimum distance. This method consists of trying many radial
velocity shifts of a template with respect to the observed spectrum and determining what velocity shift gives the best fit. 
Goodness of fit can be evaluated with a classical least-squares, but Poisson and Cash statistics are also being explored. The 
resulting radial velocity thus obtained is refined using a parabola through the points around the best fit. The 1-sigma error 
bars on the radial velocity are also determined. 

\subsection{Accuracies of the RV determinations}

The simulations with the last design of the RVS provides intrinsic accuracies of the radial velocity determinations for stars
of spectral types from K1 to B0, and magnitudes from 8.5 to 17.5. The errors range from $<$1 km/s (for bright cool stars) to
35 km/s (for faint hot stars,V=13) with 1 single transit. At the end of mission, the errors range from $<$1km/s (for bright
cool stars) to 30 km/s (for faint hot stars, V=16) with 40 transits and 3 CCDs combined. For more details about the
accuracies, we refer the reader to the technical note from Sartoretti et al. (2007) available via the GAIA website 
or on demand.


\acknowledgements
C.M. gratefully acknowledges the SOC/LOC of the GSD2008 conference for the financial assistance.
C.M. acknowledges funding from the ESA/Belgian Federal Science Policy in the 
framework of the PRODEX program (C90290).

\newpage


\begin{thebibliography}{}
\bibitem{} Chelli, A.: 2000, A\&A 358, 59
\bibitem{} Fr\'emat Y., Neiner C., Hubert A.-M., et al.: 2006, A\&A 451, 1053 
\bibitem{} Martayan C. , Hubert A.-M. , Floquet M., et al.: 2006, A\&A 445, 931
\bibitem{} Sartoretti, P.; Katz, D.; Gomboc, A.: 2007, GAIA-C6-TN-OPM-PS-006-1
\bibitem{} Skelton B.P., Waller W.H., Gelderman R.F. et al.: 1999, PASP 111, 465
\bibitem{} Staveley-Smith L., Kim S., Calabretta M.R., et al.: 2003, MNRAS 339, 87  
\bibitem{} Viala, Y.; Blomme, R.; David, M.; Delle-Luche, C.; Desert, J.-M.; Fr\'emat, Y.; Gosset, E.; Martayan, C.: 2007, GAIA-C6-SP-OPM-YV-001-3
\bibitem{} Viala, Y.; Blomme, R.; Dammerdji, Y.; Delle-Luche, C.; Fr\'emat, Y.; Gosset, E.; Jonckheere, A.; Martayan, C.: 2008, GAIA-C6-SP-OPM-YV-004-1
\end{thebibliography}
\end{document}